\documentclass[atoms,conferenceproceedings,submit,moreauthors,10pt,a4paper]{mdpi}
\preto{\abstractkeywords}{\nolinenumbers}
\firstpage{1} 
\articlenumber{1}
\pubvolume{xx}
\pubyear{2018}
\copyrightyear{2018}
\history{Received: date; Accepted: date; Published: date}
\Title{Fundamental plane of elliptical galaxies in $f(R)$ gravity: the role of luminosity}

\Author{ Vesna Borka Jovanovi\'{c} $^{1,\dagger,*}$, Predrag Jovanovi\'{c} $^{2,\dagger}$, Du\v{s}ko Borka $^{1,\dagger}$ and Salvatore Capozziello $^{3,4,5,\dagger}$}

\AuthorNames{Vesna Borka Jovanovi\'{c}, Predrag Jovanovi\'{c}, Du\v{s}ko Borka, and Salvatore Capozziello}

\address{%
$^{1}$ \quad Atomic Physics Laboratory (040), Vin\v{c}a Institute of Nuclear Sciences, University of Belgrade, P.O. Box 522, 11001 Belgrade, Serbia; vborka@vinca.rs, dusborka@vinca.rs \\
$^{2}$ \quad Astronomical Observatory, Volgina 7, P.O. Box 74, 11060 Belgrade, Serbia; pjovanovic@aob.rs \\
$^{3}$ \quad Dipartimento di Fisica "E. Pancini", Universit\`{a} di Napoli ''Federico II'', Compl. Univ. di Monte S. Angelo, Edificio G, Via Cinthia, I-80126, Napoli, Italy; 
capozzie@na.infn.it \\
$^{4}$ \quad Istituto Nazionale di Fisica Nucleare (INFN) Sez. di Napoli, Compl. Univ. di Monte S. Angelo, Edificio G, Via Cinthia, I-80126, Napoli, Italy \\
$^{5}$ \quad Gran Sasso Science Institute, Viale F. Crispi, 7, I-67100, L'Aquila, Italy }

\corres{Correspondence: vborka@vinca.rs; Tel.: +381-11-6308425}
\firstnote{These authors contributed equally to this work.}

\abstract{The global properties of elliptical galaxies are connected through the so-called fundamental plane of ellipticals, which is an empirical relation between their parameters: effective radius, central velocity dispersion and mean surface brightness within the effective radius. We investigated the relation between the parameters of the fundamental plane equation and the parameters of modified gravity potential $f(R)$. With that aim, we compared theoretical predictions for circular velocity in $f(R)$ gravity with the corresponding values from a large sample of observed elliptical galaxies. Besides, we consistently reproduced the values of coefficients of the fundamental plane equation as deduced from observations, showing that the photometric quantities like mean surface brightness are related to gravitational parameters. We show that this type of modified gravity, especially its power-law version - $R^n$, is able to reproduce the stellar dynamics in elliptical galaxies. Also, it is shown that $R^n$ gravity fits the observations very well, without need for a dark matter.}


\begin{document}

\section{Introduction}

It is well established that there are three main global observables of elliptical galaxies: the central projected velocity dispersion $\sigma_0$, the effective radius $r_e$, and the mean effective surface brightness (within $r_e$) $I_e$. It is well known that elliptical galaxies do not populate uniformly this three dimensional parameter space; they are rather confined to a narrow logarithmic plane, thus called the fundamental plane (FP) \cite{dres87,ciot97}. Any of the three parameters may be estimated from the other two. Together they describe a plane in three-dimensional space. Many characteristics of a galaxy might be correlated.

To describe the velocity of populations of stars, one defines a rotational velocity $v_c$ - net rotational velocity of a group of stars, and a dispersion $\sigma$ - the characteristic random velocity of stars. The relation $v_c/ \sigma$ characterizes the kinematics of the galaxies, and it is the main parameter which differentiates spiral from elliptical galaxies. In this manner, spiral galaxies with $v_c/ \sigma \gg 1$, are kinematically cold systems, while ellipticals with $0 < v_c/ \sigma < 1$, are kinematically hot systems.

In galactic dynamics, the Virial Theorem (VT) which relates the total mass of the galaxies with mean (rotation or dispersion) velocity of their stars, is commonly used for inferring the mass estimates of the galaxies. Stationarity is a sufficient condition for the validity of the VT, and so for ellipticals the VT holds \cite{ciot97}. In terms of $r_e$, $\sigma_0$ and $I_e$ both VT and FP can be described by the similar expressions (Eqs. (1) and (2) in \cite{busa97}). However, the values of constants $a$ and $b$ in the case of FP differ from those predicted by VT, causing the tilt of the FP with respect to the VT plain. Namely, the calculated values of tilt angle of the FP from astronomical observations give coefficients $a$ and $b$ different from those predicted by the VT ($a$ = 2, $b$ = -1). When written in logarithmic form, the two planes appear to be tilted by an angle of $\sim 15^\circ$ \cite{busa97}. This tilt can be caused by different structural and dynamical effects in elliptical galaxies \cite{ciot96}.

It is now established that the dark matter (DM) fraction is likely a major contributor to the tilt of FP, and that the DM fractions increase for larger galaxies, because the effective radii extend further out to regions dominated by the halo \cite{tara15}. However, in paper \cite{capp13}, the authors derived accurate total mass-to-light ratios $(M/L)_e$ and DM fractions, within a sphere of radius $r = r_e$ centred on the galaxies. They tested the accuracy of the mass determinations by running models with and without DM, and have found that the enclosed total $(M/L)_e$ is independent of the inclusion of a DM halo, with good accuracy and small bias.

Therefore, in this paper we try to explain tilt of the FP without DM, but using modified gravity instead. The plan of this paper is as follows. In Section 2, we briefly describe elliptical galaxies fundamental plane. We also describe used observations and methods. In Section 3 we give basic things of power-law $f(R)$ extended gravity theories in the case of a point-like source and the generalization to a spherically symmetric system which represents elliptical galaxies. In Section 4 we give a connection between the parameters of FP equation and parameters of the $R^n$ extended gravity potential. Section 5 is devoted to summary of the conclusions.

\section{Elliptical galaxies and their fundamental plane}

\subsection{Surface brightness of ellipticals}

Surface brightness $I$ is flux $F$ within angular area $\Omega^2$ on the sky ($\Omega = D/d$, where $D$ is side of a small patch in a galaxy located at a distance $d$). Let us emphasize here that $I$ is independent of distance $d$: $I = F/ \Omega^2 = L/(4 \pi d^2) \times (d/D)^2 = L/(4 \pi D^2)$, where $L$ is luminosity (see e.g. $\S$ 1.3.1 in \cite{spar07}).

Main sources of luminosity in elliptical galaxies would be: stellar plasma, hot gas, accreting black holes in the cores of stellar bulges (see e.g. \cite{spar07} and references therein). According to luminosity, their classification is the following:

\begin{enumerate}
\item Massive/luminous ellipticals ($L > 2 \times 10^{10} \, L_\odot$).
\newline \textit{These ellipticals have low central surface brightness with flat distribution (cores - regions where the surface brightness flattens).} They have lots of hot X-ray emitting gas, very old stars, lots of globular clusters, and are characterized by little rotation.
\item Intermediate mass/luminosity ellipticals ($L > 3 \times 10^9 \, L_\odot$).
\newline \textit{Their characteristic is power law central brightness distribution.} They have little cold gas, and their oblate symmetry is consistent with their moderate rotation.
\item Dwarf ellipticals ($L < 3 \times 10^9 \, L_\odot$).
\newline \textit{Their surface brightness is exponential.} There is no rotation. 
\end{enumerate}

In contrast to spirals galaxies, ellipticals show regularity in their global luminosity distributions. Surface brightness of most elliptical galaxies, measured along the major axis of a galaxy's image, can be fit by de Vaucouleurs profile: $I(r)=I_e \times 10^{-3.33 \left( (r/r_e)^{1/4}-1 \right)}$. This empirical model, also known as $r^{1/4}$ law, describes how the surface brightness varies as a function of apparent distance $r$ from the center of the galaxy. The Sersic $r^{1/n}$ profile: $I(r)=I_e \times 10^{-b_n \left( (r/r_e)^{1/n}-1 \right)}$ (the constant $b_n$ is chosen such that half of the luminosity comes from $r < r_e$), which generalizes the de Vaucouleurs profile, is also well suited to describe the surface brightness distribution of these systems (see more about this profile e.g. in \cite{ciot96,card04}). De Vaucouleurs profile is a particularly good description of the surface brightness of giant and midsized elliptical galaxies, while dwarf ellipticals are better fit by Sersic profile for $n$ = 1 (exponential profiles).

\subsection{Fundamental plane of elliptical galaxies}

The global properties of elliptical galaxies are connected, and empirical relation which shows this connection is called fundamental plane \cite{busa97}:

\begin{equation}
log(r_e) = a \ log(v_c) + b \ log(I_e) + c,
\label{equ01}
\end{equation}

\noindent with $r_e$ - effective radius (which encloses half of the total luminosity emitted by a galaxy), $v_c$ - central velocity dispersion, $I_e$ - mean surface brightness within $r_e$, and $a,b,c$ - coefficients.

Some object can be represented as a point in the parameter space $(r_e,v_c,I_e)$, and if we present it in logarithmic form, we obtain a plane \cite{bork16b,capo18}. The angle between the virial plane and FP is the so-called ''tilt''. Prediction of the VT (Virial equilibrium and constant mass-to-light $M/L$ ratio) for FP coefficients $a$ and $b$ is: $a = 2$, $b = -1$, and the empirical result (using the Virgo Cluster elliptical galaxies as a sample) gives $a = 1.4$, $b = -0.85$ \cite{bend92}. So, when presented in logarithmic form, these two planes appear to be tilted by an angle of $\sim 15^\circ$ \cite{busa97}. This can be explained by stellar population effects and by spatial non-homology in the dynamical structures of the systems. As VT uses the simplified assumptions, the tilt provides fundamental information about galaxy evolution.

\subsection{Observations and method}

The observational data of interest for our study are publicly available (in ASCI format) among the source files of the arxiv version of the paper Burstein et al. (1997) \cite{burs97}: \url{https://arxiv.org/format/astro-ph/9707037}, see 'metaplanetab1'. We use some physical properties of stellar systems: among 1150 observed galaxies, there is a sample of 401 ellipticals. In this study, we use the values from the following columns of Table 1 in \cite{burs97}: column (5) -- circular velocity (observed): $\log v_c$ (km/s); column (6) -- central velocity dispersion (derived): $\log \sigma_0$ (km/s); column (7) -- effective or half-light radius: $\log r_e$ (kpc); column (8) -- mean surface brightness within $r_e$: $\log I_e$ (L$_\odot$/pc$^2$). Here, we would like to emphasize that $\sigma_0$ is derived in that way to get the consistent values for all stellar systems, and that for elliptical galaxies the circular velocity inside effective radius is $v_c(r_e) = \sigma_0$, while for other stellar systems it is $v_c \neq \sigma_0$.

Then, using the relation for $v_c$ which consists of Newtonian contribution and the correction term due to modified gravity, we also calculate the theoretical values for circular velocity $v_c^{theor}$ and FP coefficients (see further in Section 4).

\subsection{Region of parameter space of fundamental plane}

\begin{figure}[ht!]
\centering
\includegraphics[width=0.70\textwidth]{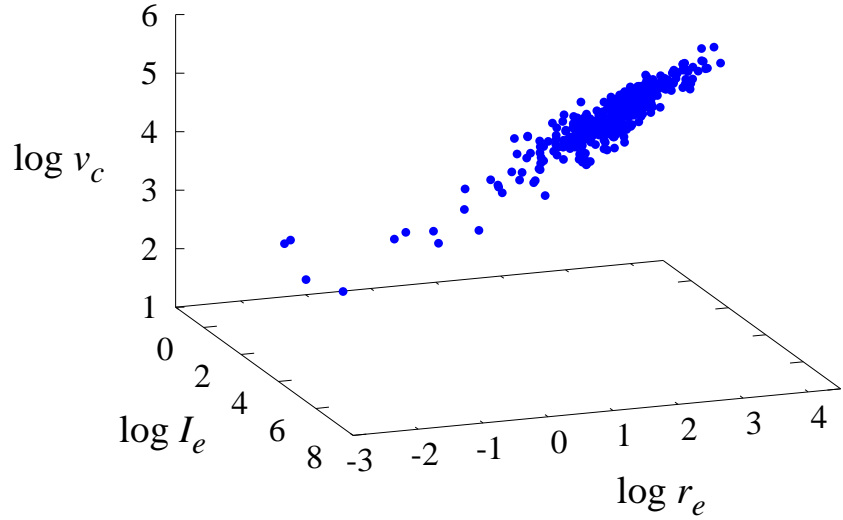}
\caption{The FP parameter space, presented by logarithms of the three parameters: mean surface brightness (within effective radius) $\log{I_e}$, effective radius $\log{r_e}$ and circular velocity $\log{v_c}$, for a sample of elliptical galaxies listed in Table 1 from \cite{burs97}. Note: in paper \cite{burs97} it is printed the first page only, and we used the whole sample of 401 ellipticals, available among the source files of its arxiv version.}
\label{fig01}
\end{figure}

Early-type galaxies are observed to populate a tight plane in the space defined by their effective radii, velocity dispersions and surface brightnesses \cite{desr07}. If elliptical galaxies were perfectly homologous stellar systems with identical stellar populations, then $r_e$, $v_c$, and $I_e$ would be related by the VT. Instead, non-homology and/or stellar population variations tend to place elliptical galaxies on a nearby fundamental plane, with the remarkably small width \cite{math03}. It has been shown that these parameters are rather stable to gravitational perturbation. The FP parameters do change during close encounters of galaxies, but within a very short time interval just before their final merger, and furthermore, the amplitudes of these changes are comparable to the scatter of the observed FP \cite{evst02}.

As we stated before, elliptical galaxies are not randomly distributed within the 3D parameter space ($I_e$, $r_e$, $v_c$), and when presented in logarithmic form, they lie in a plane. See Fig. \ref{fig01}, showing the parameter space in log scale.

\section{$f(R)$ modified gravity}

Extended Theories of Gravity (ETGs) \cite{capo11} have been proposed to explain galactic and extragalactic dynamics without introducing DM, and as such they can be used to test if FP of ellipticals could be explained taking into account only their luminous matter content. For that purpose, we adopt $f(R)$ gravity which is the straightforward generalization of Einstein's General Relativity as soon as the function is $f(R) \neq R$, that is, it is not linear in the Ricci scalar $R$ as in the Hilbert-Einstein action. As simple choice, one assumes a generic function $f(R)$ of the Ricci scalar $R$ (in particular, analytic functions) and searches for a theory of gravity having suitable behavior at all scales (at small and large scale lengths). 

We start from the action \cite{capo07}:

\begin{equation}
\mathcal{A}=\int{d^4x \sqrt{-g} \left[f(R) +\mathcal{L}_m \right]},
\label{equ02}
\end{equation}

\noindent (with $g$ - metric tensor and $\mathcal{L}_m$ - the standard matter Lagrangian), and consider power - law case:

\begin{equation}
f(R)=f_0 R^n,
\label{equ03}
\end{equation}

\noindent with $n$ the slope of the gravity Lagrangian, and $f_0$ a dimensional constant (dimensions for $f_0$ chosen in such
a way to give $f(R)$ the right physical dimensions).

$R^n$ gravity is the power-law version of $f(R)$ modified gravity. In the weak field limit, its potential (generated by a pointlike mass $m$ at the distance $r$) is \cite{capo07}:
\begin{equation}
\Phi(r) = - \frac{G m}{2 r} \left[ 1 + \left( \frac{r}{r_c} \right)^{\beta} \right],
\label{equ04}
\end{equation}

\noindent with $r_c$ - scalelength depending on the gravitating system properties, and $\beta$ - universal constant:

\begin{equation}
\beta = \frac{12n^2 - 7n - 1 - \sqrt{36n^4 + 12n^3 - 83n^2 + 50n + 1}}{6n^2 - 4n + 2}.
\label{equ05}
\end{equation}

\noindent Here, the case $n = 1 \Rightarrow \beta = 0$ represents Newtonian case. About the power-law fourth-order theories of gravity, as well as about determination of the space parameters of $f(R)$ gravity, see \cite{capo07,bork12,bork13,zakh14,bork15,bork16a,capo17}.

The solution (\ref{equ04}) has been obtained in the case of a point-like source, but it can be generalized to the case of extended systems. The generalization of Eq. (\ref{equ04}) to a spherically symmetric system, gives the correction term of the potential \cite{capo07}:

\begin{equation}
\Phi_c(r) = - \frac{\pi G \alpha r_c^2}{3} \left [ {\cal{I}}_1(r)
+ {\cal{I}}_2(r) \right ]
\label{equ06}
\end{equation}

\noindent with parameter $\alpha = 1$ for $R^n$ gravity, and with \,:

\begin{eqnarray}
{\cal{I}}_1 = 3 \pi \int_{0}^{\infty}{(\xi^2 + \xi'^2)^{(\beta
- 1)/2} \rho(\xi') \xi'^2 d\xi'}
~{\times}~ {_2F_1}\left [\left \{ \frac{1 - \beta}{4}, \frac{3
- \beta}{4} \right\}, \{2\}, \frac{4 \xi^2 \xi'^2}{(\xi^2 +
\xi'^2)^2} \right ] \ ,
\label{equ07}
\end{eqnarray}

\begin{eqnarray}
{\cal{I}}_2 = 4 (1 - \beta) \xi \int_{0}^{\infty}{ (\xi^2 +
\xi'^2)^{(\beta - 3)/2} \rho(\xi') \xi'^2 d\xi'}
~{\times} \ {_3F_2}\left [\left \{ 1, \frac{3 - \beta}{4},
\frac{5 - \beta}{4} \right \}, \left \{ \frac{3}{2}, \frac{5}{2}
\right \}, \frac{4 \xi^2 \xi'^2}{(\xi^2 + \xi'^2)^2} \right ]  \ ,
\label{equ08}
\end{eqnarray}

\noindent where $\xi$ is generically defined as $\xi = r/r_c$, and the notation for the hypergeometric functions is used: ${_pF_q}[\{a_1, \ldots, a_p\}, \{b_1, \ldots, b_q\}, x]$.

\section{Fundamental plane in $f(R)$ gravity}

\subsection{Recovering fundamental plane from $R^n$ gravity}

We want to show the connection of the FP of elliptical galaxies with $R^n$ gravity potential, by showing the correlation between the corresponding parameters:
\begin{itemize}
\item[-] addend with $r_e$: correlation between $r_e$ and $r_c$;
\item[-] addend with $\sigma_0$: correlation between $\sigma_0$ and $v_{vir}$ ($v_{vir}$ - virial velocity);
\item[-] addend with $I_e$: correlation between $I_e$ and $r_e$ (through $r_c/r_e$ ratio).
\end{itemize}

Here, the reader should note that $r_e$ is the \underline{observational} gravitational radius (derived from photometry, i.e. its value is determined by the self-gravitating luminous matter content in the inner part of the elliptical galaxy), and $r_c$ is the \underline{theoretical} gravitational radius (from $R^n$ gravity). We assumed that these two radii were mutually proportional and we tested their different ratios. It is also important here to emphasize that if we introduce the assumption $r_c \sim r_e$, then in point $r_e$ the integrals $\mathcal{I}_1(r_e)$ and $\mathcal{I}_2(r_e)$ (see Eqs. (\ref{equ07}) and (\ref{equ08})) do not depent on $r$, then $\Phi_c(r_e)$ (see Eq. (\ref{equ06})) does not depend on $r$, so the correction velocity is $v_{c,corr}(r_e) = 0$. In other words, under the condition $r_c \sim r_e$, $R^n$ gravity gives the same $\sigma_0$ for elliptical galaxies as in Newtonian case. For more details about our method, see \cite{bork16a,capo18}.

\begin{figure}[ht!]
\centering
\includegraphics[width=0.95\textwidth]{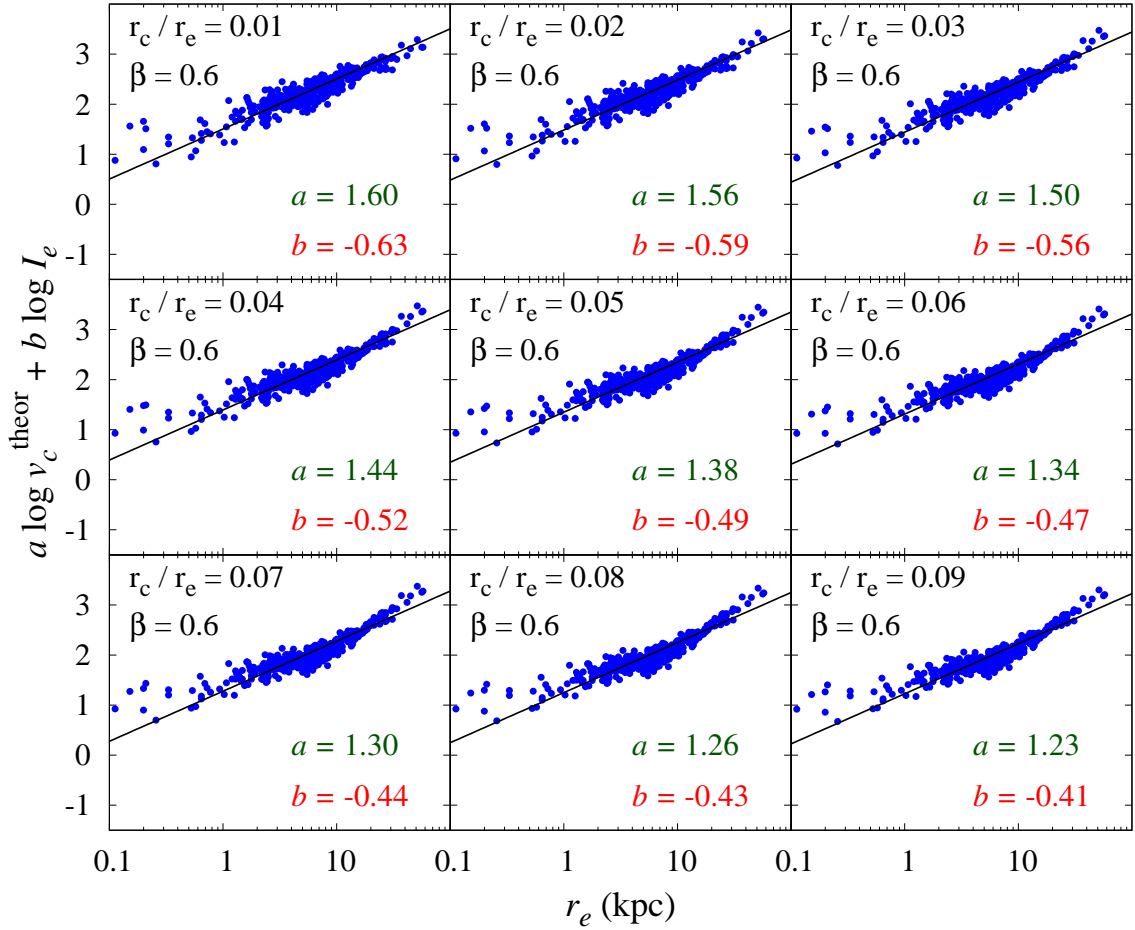}
\caption{Fundamental plane of elliptical galaxies with calculated circular velocity $v_c^{theor}$, observed effective radius $r_e$ 
and observed mean surface brightness (within the effective radius) $I_e$. For each given pair of $R^n$ gravity parameters ($r_c$, 
$\beta$), i.e. for the certain cases $r_c / r_e$ = 0.01, 0.02, 0.03, 0.04, 0.05, 0.06, 0.07, 0.08 and 0.09, and $\beta$ = 0.6, we 
present calculated FP coefficients ($a$, $b$). Black solid line is result of 3D fit of FP.}
\label{fig02}
\end{figure}

\subsection{Fundamental plane coefficients}

The empirical result for FP coefficients are given in Bender et al. 1992 \cite{bend92}: $a = 1.4$, $b = -0.85$. Empirically derived values means that $a$ and $b$ are calculated using observed FP parameters, as coefficients of the FP equation (with the Virgo Cluster elliptical galaxies as a sample). The test for our method is recovering this profile: starting from the gravitational potential derived from $f(R)$ gravity, these values have to be consistently reproduced.

According to Eq. (25) for the rotation curve $v_c(r)$ in paper \cite{capo07}, it consists of Newtonian part and the correction term due to $R^n$ gravity:

\begin{equation}
v_c^2\left( r \right) = \frac{v_{c,N}^2\left( r \right)}{2} + 
\frac{r}{2} \frac{\partial {\Phi_c}}{\partial r},
\label{equ09}
\end{equation}

\noindent and therefore we used the above expression to calculate $v_c^{theor}$ - the theoretical prediction of $R^n$ gravity for circular velocity $v_c$, in the case of extended spherically symmetric systems, taking into account the so called Hernquist profile for density distribution \cite{hern90}:

\begin{equation}
\rho(r) = \dfrac{a_H M}{2 \pi r (r + a_H)^3}, \ \ \ a_H = 
\dfrac{r_e}{1 + \sqrt{2}}.
\label{equ10}
\end{equation}

FP of elliptical galaxies with 3D fit, with the calculated values $v_c^{theor}$, and the observed values $r_e$, $I_e$, presenting the dependence of FP parameters $(a,b)$ on parameters of $R^n$ gravity $(r_c,\beta)$, we show in Fig. \ref{fig02}. In this figure we presented only the case $\beta = 0.6$, but we tested other values of this parameter in the similar way as well. The phrase ''3D fit'' denotes a fit of a function $z$ depending on two independent variables $(x,y)$ to the observational data, and in this case Eq. (\ref{equ01}) is fitted with function $z(x,y) = ax + by + c$, where $x = \log(v_c^{theor})$, $y = \log(I_e)$ and $z = \log(r_e)$, using the least-squares algorithm implemented in ''fit'' command of Gnuplot (\url{http://www.gnuplot.info/}). As a result we obtained the best fit coefficients of FP equation: $a$, $b$ and $c$. The procedure is the following (see our Ref. \cite{bork16a} for a detailed explanation): we varied $R^n$ gravity parameters $(r_c, \beta)$ and for each given pair of the parameters $(r_c, \beta)$, i.e. for the certain ratios $r_c / r_e$ and certain values $\beta$, we calculated the terms $x$, $y$ and $z$ and finally obtained coefficients $(a,b,c)$. Once this procedure is performed, the obtained values of $a$ and $b$, are compared with $a$ and $b$ values obtained from observations \cite{bend92}. As it can be seen from Fig. \ref{fig02}, a smaller $r_c/r_e$ ratio results with a larger value for FP parameter $a$ and smaller value of parameter $b$, obtained by fitting the FP equation through the $(I_e,r_e,v_c^{theor})$ data points. The best fit (and the smallest scatter of these data points) is obtained for $r_c/r_e \approx 0.05$.

\subsection{Luminosity and parameters of $R^n$ gravity}

Correlation of $I_e$ with $r_c$ is reflected through the coefficient $b$ of FP in equation: $r_e \sim I_e^b \times v_c^a$ \cite{bend92,bend93}. This means that being $r_c$ related to $r_e$ through $r_c \sim r_e$, also $r_c$ is related to $I_e$. On the other hand, analytic expression for $v_c$ includes both modified gravity parameters ($r_c$,$\beta$).

Coefficients $a,b,c$ of FP are also correlated with ($r_c$,$\beta$): see Table 3 in our paper \cite{bork16a}.

In general, this means that photometric quantities like $I_e$ are related (in a complex way) to the parameters of modified gravitational potential.

\section{Discussion and conclusions}

Here we studied possible connection between the empirical parameters of FP for ellipticals and the theoretical parameters of $R^n$ gravity in order to test if corrections predicted by this type of gravity could explain both photometry and dynamics of ellipticals without DM hypothesis.

Our main conclusions may be summarized as follows:

\begin{itemize}

\item We connected fundamental plane of elliptical galaxies with $R^n$ gravity potential, relating together observational and theoretical quantities (i.e. tying the corresponding FP and $R^n$ parameters).
\item We reproduced the FP generated by the power law $f(R)$ gravity without considering the presence of DM in galaxies.
\item We obtained that the characteristic radius $r_c$ of $R^n$ gravity is proportional to the effective radius $r_e$: more precise, $r_c \approx 0.05 r_e$ gives the best fit with data. This fact points out that the gravitational corrections induced by $R^n$ can lead photometry and dynamics of the system.
\item We demonstrated that not only stellar kinematics of ellipticals could be affected by modified gravity (as we have already shown in our previous papers), but also their most important physical properties, such as their luminosity.
\end{itemize}

We compared, for the first time, theoretical predictions for circular velocity in $f(R)$ gravity with the corresponding values from the large sample of observed elliptical galaxies. Using gravitational potential derived from $f(R)$ gravity, we consistently reproduced the values of FP parameters. We pointed out that the photometric quantities, like mean surface brightness, are related to gravitational parameters. Also, we explained that $R^n$ gravity fits the observations very well, taking into account only the luminous matter content of ellipticals, hence not needing DM.

\acknowledgments{This work is supported by Istituto Nazionale di Fisica Nucleare, Sezione di Napoli, Italy, iniziative specifiche TEONGRAV and QGSKY, and by Ministry of Education, Science and Technological Development of the Republic of Serbia, through the project 176003 ''Gravitation and the Large Scale Structure of the Universe''. The authors also acknowledge the support by Bilateral cooperation between Serbia and Italy 451-03-01231/2015-09/1 ''Testing Extended Theories of Gravity at different astrophysical scales'' and of the COST Action CA15117 (CANTATA), supported by COST (European Cooperation in Science and Technology). The authors would like to thank Dr. Vladimir Rekovi\'{c} for improving the English of the paper.}

\authorcontributions{All the coauthors participated in calculation and discussion of obtained results.}

\conflictsofinterest{The authors declare no conflict of interest.}

\abbreviations{The following abbreviations are used in this manuscript:\\

\noindent 
\begin{tabular}{@{}ll}
DM & Dark matter \\
ETGs & Extended Theories of Gravity \\
FP & Fundamental plane \\
GC & Galactic Center \\
GR & General Relativity \\
VT & Virial Theorem \\

\end{tabular}}

\reftitle{References}

\end{document}